\documentclass[twocolumn,showpacs,preprintnumbers,amsmath,amssymb]{revtex4}

\usepackage{epsfig}
\usepackage{graphicx}
\usepackage{dcolumn}

\newcommand{\beq}{\begin{equation}}
\newcommand{\eeq}{\end{equation}}
\newcommand{\beqa}{\begin{eqnarray}}
\newcommand{\eeqa}{\end{eqnarray}}
\newcommand{\lam}{\lambda}

\newcommand{\rh}{\rho}

\newcommand{\al}{\alpha}
\newcommand{\si}{\sigma}

\newcommand{\om}{\omega}

\newcommand{\la}{\langle}
\newcommand{\ra}{\rangle}

\def\jpb#1{{ J.\ Phys.\ B} {\bf#1}}

\def\pra#1{{ Phys.\ Rev. A\/} {\bf#1}}
\def\prb#1{{ Phys.\ Rev. B\/} {\bf#1}}

\def\prl#1{{ Phys.\ Rev.\ Lett.} {\bf#1}}

\begin{document}

\title{Pairwise Concurrence Dynamics:  A Four-Qubit Model}

\author{M.\  Y\"ona\c{c}, Ting Yu\footnote{Email: ting@pas.rochester.edu} and J.\ H.\  Eberly\footnote{Email: eberly@pas.rochester.edu}}
\affiliation{ Rochester Theory Center
for Optical Science and Engineering, and  Department of Physics and
Astronomy, University of Rochester,  New York 14627, USA }


\date{February 14, 2006}

\begin{abstract}
We examine entanglement dynamics via concurrence among four two-state systems labeled $A, ~a, ~B, ~b$. The four systems are
arranged on an addressable ``lattice" in such a way that $A$ and $a$ at one location labeled $Aa$ can interact with each other via
excitation exchange, and the same for $B$ and $b$ at location $Bb$. The $Aa$ location is prepared entangled with the $Bb$ location,
but their mutual complete isolation prevents interaction in the interval between actions of an external addressing agent. There are six pairwise
concurrences on the lattice, and we follow their evolution in the interval between external actions. We show how entanglement evolves and may exhibit the non-analytic effect termed  entanglement sudden death (ESD), with periodic recovery. These loss and gain processes may be
interpreted as entanglement transfer between the subsystems.
\end{abstract}


\maketitle

\maketitle
\section{Introduction}
\label{modelsection}

One of the challenges of realizing quantum information processing is control of evolution of qubits in the presence of environmental noises
and manipulation inaccuracies \cite{MC}.  This control may be easily achieved for a single qubit \cite{Slichter78},  but for many qubits, entanglement dynamics
has been a difficult subject and has attracted extensive interest recently ranging from two-qubit systems \cite{Yu-Eberly02,Yu-Eberly03,Yu-Eberly04,Yu-Eberly05,Yu-Eberly06B,Lidar04,Jakobczyk-Jamroz04,Tolkunov-etal05,Ficek-Tanas06}, to continuous variables \cite{Halliwell,Ban06}, spin systems \cite{Zyczkowski02,Diosi03,Solenov05},
and multi-partite systems \cite{Carvalho-etal04,Carvalho-etal05,phys,Lu,kuzi0,kuzi1,kuri2,sun}.  Moreover,  in  \cite{Santos-etal06,nature}, proposals have been made for the direct measurement of finite-time disentanglement in cavity QED, and real-time detection of entanglement sudden death (ESD) has been reported very recently \cite{Davi}.

The behavior of qubits in quantum computing environments is expected to be determined by external operations, addressed to specific qubit locations, that manipulate transitions (operate gates) in sequences determined by quantum algorithms. Within a qubit network at different locations there may be lengthy intervals between such externally mandated gate operations, and the small ``network" we examine here is intended to provide insight as to the preservation or degradation of entanglement between remote qubits on the lattice in these intervals. Our network has a dual atom-photon basis in the sense that each active lattice site incorporates a photonic and a non-photonic element that communicate with each other by an emission-absorption interaction. A cavity QED interpretation is possible \cite{Yu-Eberly04,Yonac}.

One should not expect changes in entanglement between two arbitrary units to be determined only by their mutual interaction strength.
Results obtained below demonstrate changes in the entanglement of remote units of the lattice that have zero mutual interaction
and no communication, quantum or classical.

We begin by specifying the interactions that will be allowed. These
are intended to permit systems $A$ and $a$ to exchange a degree of
excitation (a spin flip or a photon, for example). The same should
be true of $B$ and $b$. Thinking in the cavity QED context, a suitable
beginning would be a sum of two separate Jaynes-Cummings
Hamiltonians (with $\hbar = 1$) \cite{Yonac,JC,Eberlyetal80}:
\beqa
H_{\rm tot} &=& \frac{\om_0}{2}\sigma_z^A  +
g(a^{\dagger}\sigma_{-}^A + \sigma_{+}^A a)  + \om a^{\dagger}a \nonumber\\
 && + \frac{\om_0}{2}\sigma_z^B + g(b^{\dagger}\sigma_{-}^B +
 \sigma_{+}^B b) + \om b^{\dagger}b,
\eeqa
where $\omega_0$ and $\omega$ are the frequencies of atoms and cavities, respectively,  and $g$ is the coupling constant between the atoms and their cavities. Physical realizations of our scenario does not appear out of the question, as the Jaynes-Cummings model has been realized in the laboratory in several well-known ways \cite{Harochegroup, Winelandgroup, Kimblegroup}.

Clearly the remoteness of site $Aa$ from site $Bb$  is perfect in
this case, since there can be no interaction between the sites at
all with the Hamiltonian above, although a familiar type of
interaction will take place on each site. It is not difficult to
imagine the types of external probes that could be coupled to the
sites, but we will be interested in evolution of entanglement in a
time interval between such external controls.

The eigenstates of this Hamiltonian are products of the eigenstates
(dressed states) of the separate JC systems, which are well known
\cite{JC}. We will write for either $Aa$ or $Bb$ \beq
\label{JCEvalEqn} H_{\rm JC}|\psi_n^{\pm}\ra = \lambda_n^{\pm}
|\psi_n^{\pm}\ra. \eeq
We will think of the non-photonic
(atomic or spin) states as ``excited" and ``ground" states denoted
by $|e\ra$ and $|g\ra$, and denote the cavity mode's photon states
by the photon number $n$. Then the single-atom eigenvalues are given
by:
\beq \label{JCEvals} \lambda_n^{\pm} = n\omega +
\frac{1}{2}\Big(\Delta \pm \sqrt{\Delta^2 + G_n^2}\Big), \eeq and
the dressed states are these:
\beqa
\label{JCstates}
|\psi_0\ra &=& |g;0\ra \\
|\psi_n^+\ra &=& c_n|e; n-1\ra + s_n|g; n\ra\,\,\, (n>0)\\
|\psi_n^-\ra &=& -s_n|e; n-1\ra + |c_n|g; n\ra\,\,\, (n>0).
\eeqa
In
these equations we have introduced some convenient abbreviations:
\beq
\label{cands}
c_n \equiv \cos(\theta_n/2) \quad {\rm and}\quad s_n \equiv
\sin(\theta_n/2),
\eeq
where the rotation angle $\theta_n$ can be
identified with the Bloch sphere polar angle and is defined in the
usual way:
\beq \label{thetadef}
\cos\theta_n \equiv
\frac{\Delta}{\sqrt{\Delta^2 + G_n^2}} \quad {\rm and} \quad
\sin\theta_n \equiv \frac{G_n}{\sqrt{\Delta^2 + G_n^2}},
\eeq
where $\Delta = \omega- \om_0$ is the detuning and
\beq
G_n = 2g\sqrt{n}.
\eeq
In order to achieve a four-qubit model, we restrict the cavities to a flip-flip operation. In the examples to be treated we will need the true ground state $|\psi_0\ra = |g;0\ra $ and the two dressed states for $n=1$. These three states are closed under the JC Hamiltonian for each site.  In other words,  we use only $n=1$ in the equations above, so the subscript $n$ can mostly be ignored and we will frequently drop it ($\lam_n \to \lam$, $c_n \to c$, etc.).

The paper is organized as follows:  After a brief introduction to Concurrence chosen as the measure of entanglement in Sec. ~\ref{concurrence/standard}, the evolution of two important sets of entangled pure states toward decoherence is discussed In Sec. \ref{decoherence}. We examine evolution of concurrence between all qubit pairs in this four-qubit model. In particular, we show the existence of parameter zones where entanglement sudden death (ESD) takes place periodically. Moreover, we demonstrate a case where  the loss of atom-atom entanglement is compensated by concurrence gain in the photon-photon space. This loss and gain may be interpreted as  entanglement transfer between atomic and photonic variables. The interconnections between pairwise concurrences in this four-partite system are discussed in detail.  We conclude in Sec. \ref{conclusion}.

\section{Measure of Entanglement}
\label{concurrence/standard}

The JC dressed states are already atom-photon entangled states, and this entanglement has interesting consequences if the cavities are fairly highly excited, as Gea-Banacloche originally pointed out \cite{Banacloche},  but this is a one-site entanglement arising simply from the allowed interaction. Here we are only interested in entanglement as a non-local phenomenon that bridges any remote separation of sites on the lattice.

In the general context of entanglement we note that there is no accepted and practically workable criterion for measuring entanglement (for determining separability) of arbitrary four-particle states. Our purposes will be satisfied by working with two-particle mixed states obtained from the time-evolving four-particle pure state, since the two-qubit domain of entanglement is where all familiar measures agree. That is, entropy of formation, Schmidt number, tangle, negativity, and concurrence are numerically somewhat different, but in the two-qubit domains of their applicability they are in full agreement when they signal complete separability, i.e., the lack of entanglement.

We will adopt Wootters' concurrence \cite{Wootters} as our measure
in this discussion, mainly for its relevance for mixed states and
the  convenience of its definition and normalization:
\beq \label{definationc}
C(\rh) = \max\{0,\sqrt{\lam_1} - \sqrt{\lam_2} - \sqrt{\lam_3} - \sqrt{\lam_4} \},
\eeq
where the quantities $\lam_i$ are the eigenvalues in decreasing order of the matrix
\beq \zeta=\rho(\sigma_y\otimes
\sigma_y)\rho^*(\sigma_y\otimes \sigma_y),
\label{concurrence}
\eeq
where $\rh^*$ denotes the complex
conjugation of $\rh$ in the standard basis  and
$\si_y$ is the Pauli matrix expressed in the same basis as:
\begin{equation}
\si_y= \left(\begin{array}{clcr}
0  &  -i\\
i  &   0 \\
\end{array}
       \right).
\end{equation}
Notice that  $1 \ge C \ge 0$, where $C=0$ indicates zero entanglement and $C=1$ means maximal pure state entanglement, as in a Bell state, for example.

It will turn out that several universal features will appear in the entanglements arising from our four-qubit model. The simplest ones are these. First, all reductions to two-qubit form, obtained by tracing over the other two qubits, will yield a two-qubit mixed state always having the ``X" form \cite{ye1}:
\beq
\label{mixedstate0}
\rho = \left[
\begin{array}{cccc}
a & 0 & 0 & z \\
0 & b & 0 & 0 \\
0 & 0 & c & 0 \\
z^* & 0 & 0 & d
\end{array} \right], \\
\eeq
where $a+b+c+d = 1$. Second, since the concurrence of this mixed state is easily found to be
\beq \label{generalC}
C = 2\max\{0, |z| - \sqrt{ad}\} \equiv 2\max\{0, Q\},
\eeq
it is clear that $Q$, defined as
\beq \label{Qdefined}
Q \equiv |z| - \sqrt{ad},
\eeq
will be an important quantity. We will mention at the end certain ``conservation" properties that derive from $Q$ in some cases because it can be negative, whereas $C$ cannot.

\section{Pairwise entanglement dynamics of qubits}
\label{decoherence}

The entanglement dynamics of  qubits is generally very complex even for the case of two qubits, as  decoherence  inevitably evolves a pure state into a mixed state.  It is easy to see that for an arbitrary initial state of the two-atom system there is no exact solution to allow  a clear-cut analysis. However, we find that two sets of pure initial states for two atoms $A, B$  can be examined exactly in our model. These initial states are superpositions of the two types of Bell states: $|\Phi^{\pm)}\ra \sim |e_A, e_B\ra \pm |g_A, g_B\ra $ and $|\Psi^{\pm)}\ra \sim |e_A, g_B\ra \pm |g_A, e_B\ra $ . Thus we denote  superpositions within each type as follows:
 \beq
\label{PhiZero1}
|\Phi_{AB}\ra =\cos\alpha|e_A, e_B\ra + \sin\alpha|g_A, g_B\ra,
\eeq
and
\beq \label{PZero1}
|\Psi_{AB}\ra = \cos\alpha|e_A, g_B\ra + \sin\alpha|g_A, e_B\ra.
\eeq
These initial states are obviously interesting and physically relevant.  It is easy to see that  $\alpha=\pi/4$ reproduces the standard Bell states.  It should be noted for certain types of initial mixed states such as X-states \cite{ye1}, explicit results can also be obtained.

In the two-site situation under consideration there are, in principle, six different concurrences that can provide information about the bipartite entanglements that may arise. With an obvious notation we can denote these as $C^{AB}, ~C^{ab}, ~C^{Aa}, ~C^{Bb}, ~C^{Ab}, ~C^{Ba}$. All except $C^{Aa}$ and $C^{Bb}$ report remote entanglements across the $(Aa)$-$(Bb)$ gap, and we will be dealing with these four cases. That is, in what follows, we will examine in detail the entanglement evolution between two atoms, two cavities or between one atom and the opposite cavity.

\subsection{Partially entangled Bell states  $|\Phi_{AB}\ra$}
\label{firstpair}

In this section, we will take a combination of the $|\Phi^{\pm}\ra$ Bell states as our initial state for each of two atoms labeled $A$ and $B$, located without communication in  two separate cavities with modes labeled $a$ and $b$. The chosen total initial state for the atoms plus cavities is:
\beqa
\label{PZero}
|\Phi(0)\ra & = &
|\Phi_{AB}\ra \otimes |0_a,0_b\ra \\
&=&
(\cos\alpha|e_A, e_B\ra + \sin\alpha|g_A, g_B\ra) \otimes |0_a,0_b\ra. \nonumber \eeqa

To prepare for the time evolution we express these states in terms of the dressed eigenstates given below (\ref{JCstates}), and obtain:
\beqa \label{inverseJCstates}
|e_A, 0_a\ra &=& c|\psi_1^+\ra - s|\psi_1^-\ra\nonumber\\
|g_A, 1_a\ra &=& s|\psi_1^+\ra + c|\psi_1^-\ra \quad {\rm
and}\nonumber \\
|g_A, 0_a\ra &=& |\psi_0\ra.
\eeqa
Thus the initial atom-atom entangled state is
\beqa
&&|\Phi(0)\ra = \cos\alpha|e_A,0_a\ra\otimes|e_B,0_b\ra \nonumber \\
&& + \sin\alpha |g_A, 0_a\ra \otimes |g_B,0_b\ra \nonumber\\
&& = \cos\alpha (c|\psi_1^+\ra_A - s|\psi_1^-\ra_A) \otimes  (c|\psi_1^+\ra_B \nonumber \\
&&- s|\psi_1^-\ra_B) + \sin\alpha|\psi^0\ra_A \otimes|\psi^0\ra_B.
\eeqa

Evolution in time is now easily arranged because the evolution of
the dressed states is immediate:
\beq \label{psiEvolution}
|\psi^{\pm}(t)\ra = e^{-i\lambda^{\pm}t}~|\psi^{\pm}(0)\ra. \eeq
Since the combination of coefficients in our $|\Psi(0)\ra$  uniquely
associates $c$ with $|\psi^+\ra$ and $s$ with $|\psi^-\ra$, the time
evolution can be transferred to the $c$ and $s$ symbols. We will henceforth consider them carrying the time-evolution exponents. We will use the notation $c_0$ and $s_0$ to refer to their values at $t=0$ (no relation to the $n=0$ subscripts in Eq.~(\ref{cands}). Then we can write (temporarily indicating explicit time dependences for the $c$'s):
\beqa |\Phi(t)\ra &=& \cos\alpha \Big(c(t) |\psi_1^+\ra_A
- s(t)|\psi_1^-\ra_A \Big) \nonumber \\
& \otimes & \Big(c(t)|\psi_1^+\ra_B - s(t)|\psi_1^-\ra_B\Big) \nonumber \\
&+& \sin\alpha|\psi^0\ra_A \otimes|\psi^0\ra_B,
\eeqa
where the $|\psi^{\pm}\ra$ states will continue to refer to the states at $t=0$.

In order to take traces over individual atoms or cavities we need to
revert to the bare bases and this leads to:
\beqa
\label{psifunction}
|\Phi(t)\ra &=& \cos\alpha \Big(c(t)(c_0|e_A,0_a\ra + s_0|g_A,1_a\ra) \nonumber\\
&-& s(t)(-s_0|e_A,0_a + c_0|g_A, 1_a\ra)\Big)
\otimes \Big( c(t)(c_0|e_B,0_b\ra \nonumber \\
&+& s_0|g_B,1_b\ra) - s(t)(-s_0|e_B,0_b + c_0|g_B, 1_b\ra)\Big) \nonumber \\
&+& \sin\alpha|g_A, 0_a\ra \otimes |g_B, 0_b\ra \nonumber \\
&=& \cos\alpha\Big((cc_0 +ss_0)|e_A,0_a\ra + (cs_0 -sc_0)|g_A, 1_a\ra \Big) \nonumber \\
&\otimes& \Big((cc_0 +ss_0)|e_B,0_b\ra + (cs_0 -sc_0)|g_B, 1_b\ra \Big) \nonumber\\
&+& \sin\alpha |g_A,0_a\ra \otimes |g_B,0_b\ra. \eeqa

For orientation we will first focus on just $C^{AB}$, which has been
noted previously \cite{Yonac}. This requires that we trace out $a$
and $b$ photon modes, in order to get the two-qubit mixed state needed for calculation of $AB$ concurrence. The projections that are needed are:
\beqa
\la 0_a, 0_b|\Phi(t)\ra &=& \cos\alpha(cc_0 + ss_0)^2|e_a,e_B\ra + \sin\alpha |g_A, g_B\ra \nonumber \\
\la 1_a,0_b|\Phi(t)\ra &=& \cos\alpha (cs_0 - sc_0)(cc_0 + ss_0) |g_A,e_B\ra \nonumber \\
\la 0_a,1_b|\Phi(t)\ra &=& \cos\alpha (cc_0 + ss_0)(cs_0 - sc_0)
|e_A,g_B\ra\nonumber\\
\la 1_a,1_b|\Phi(t)\ra &=&0.
\eeqa

These are simple enough to see that the $AB$ mixed state has the form mentioned in the previous Section:
\beq
\label{mixedstate}
\rho^{AB} = \left[
\begin{array}{cccc}
a & 0 & 0 & z \\
0 & b & 0 & 0 \\
0 & 0 & c & 0 \\
z^* & 0 & 0 & d
\end{array} \right], \\
\eeq
for which the concurrence has the stated form
\beq \label{concurrenceEqn}
C^{AB} = 2 ~max\{0, |z| - \sqrt{bc}\}.
\eeq

We easily find the following \beqa
|z|^2 &=& \sin^2\alpha\cos^2\alpha |(cc_0 + ss_0)|^2 \nonumber \\
&=& \sin^2\alpha\cos^2\alpha (c_0^4 + s_0^4 +2c_0^2s_0^2\cos\delta t)^2, \nonumber \\
|z| &=& \sin\alpha \cos\alpha(c_0^4 + s_0^4 +2c_0^2s_0^2\cos\delta t), \nonumber \\
b = c &=& \cos^2\alpha |cc_0 + ss_0|^2~|cs_0 - sc_0|^2 \nonumber \\
&=& \cos^2\alpha (c_0^4 + s_0^4 +2c_0^2s_0^2\cos\delta t) \nonumber \\
& \times & c_0^2 s_0^2 (2-2\cos\delta t) . \eeqa

For simplicity we will evaluate this in the resonance case,
$\theta_n = \pi/2$, where $c_0 = s_0 = 1/\sqrt{2}$. Then we find
\beq |z| - \sqrt{bc} = \frac{1}{4}\cos^2\alpha (2+2\cos\delta
t)[\tan\alpha - \sin^2(\delta t/2)], \eeq where $\delta|_{\Delta =0}
= G,$ so the expression for concurrence turns out to be: \beq
\label{e.CAB-ESD} C^{AB} = 2\max\{0, Q^{AB} \}, \eeq where $Q^{AB} =
\cos^2\alpha\cos^2(Gt/2)[\tan\alpha - \sin^2(Gt/2)]$.

\begin{figure}[!h]
\epsfig{file=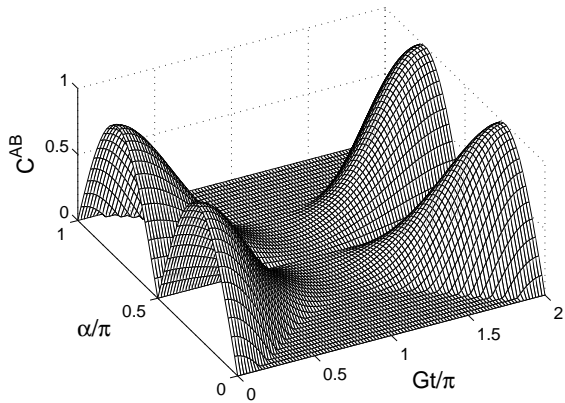, width=8cm} \caption{{\footnotesize
\label{fig1} Surface plot of the concurrence $C^{AB}$ for the $|\Phi_{AB}\ra$ type of initial state, as a function of time and the parameter $\alpha$, from Eq. (\ref{e.CAB-ESD}).
}}\end{figure}

\begin{figure}[!h]
\epsfig{file=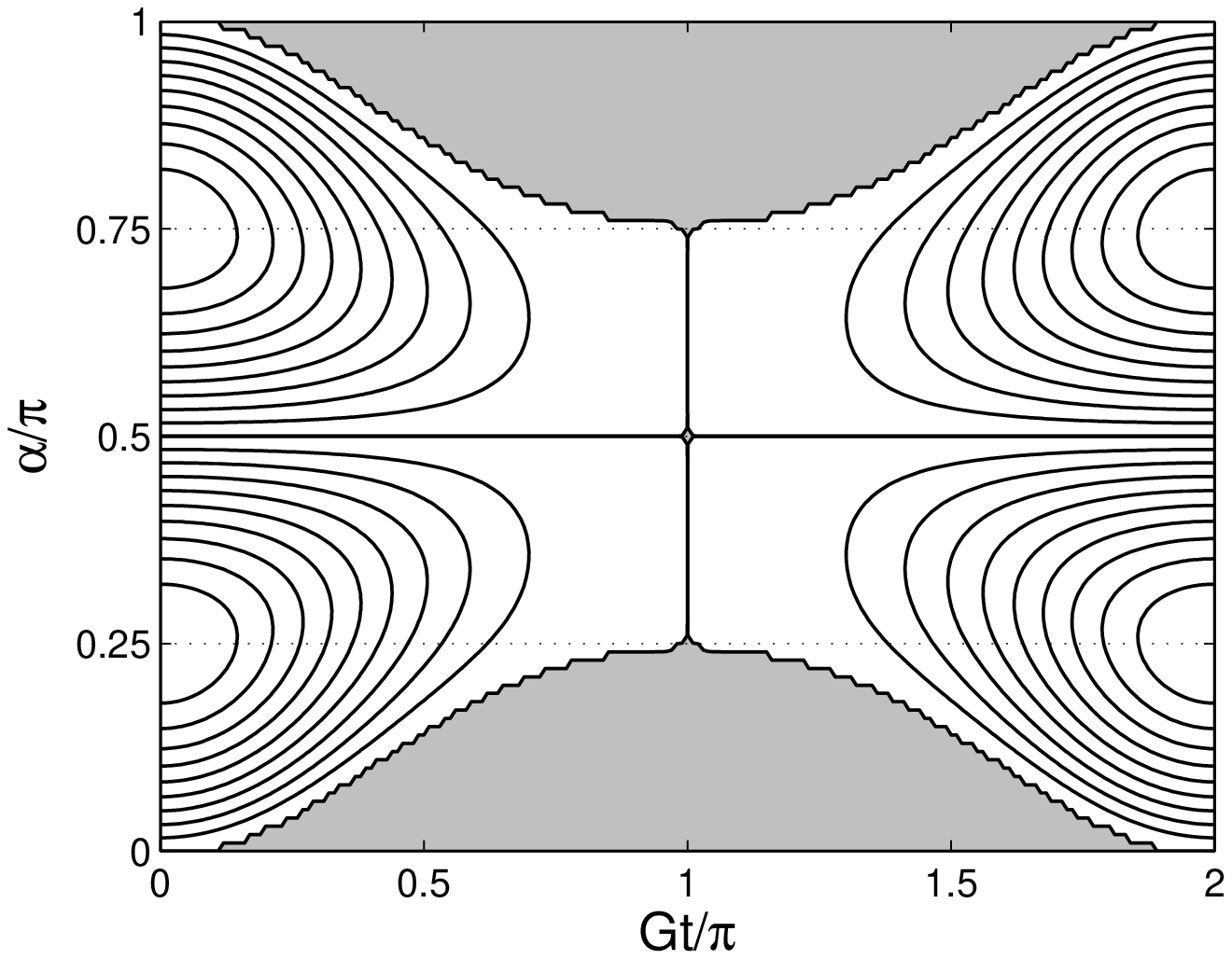, width=8.0 cm}
\caption{{\footnotesize \label{fig2}  Contour plot of the
concurrence $C^{AB}$ for the $|\Phi_{AB}\ra$ type of initial state as a
function of time and parameter $\alpha$. The innermost contours indicate a concurrence value of 0.9, and the value drops by $\Delta_c = 0.1$ between two consecutive contours. Regions of sudden death are painted in
gray.}}\end{figure}

\begin{figure}[!t]
\epsfig{file=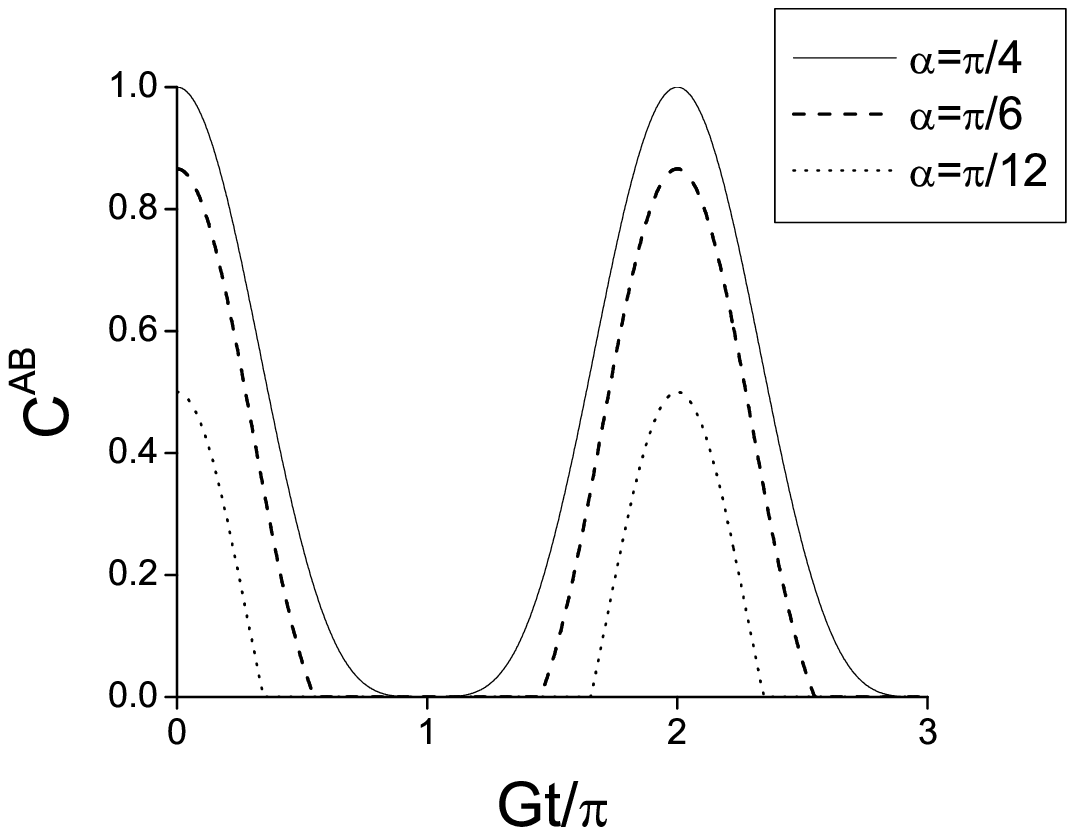, width=6cm}
\epsfig{file=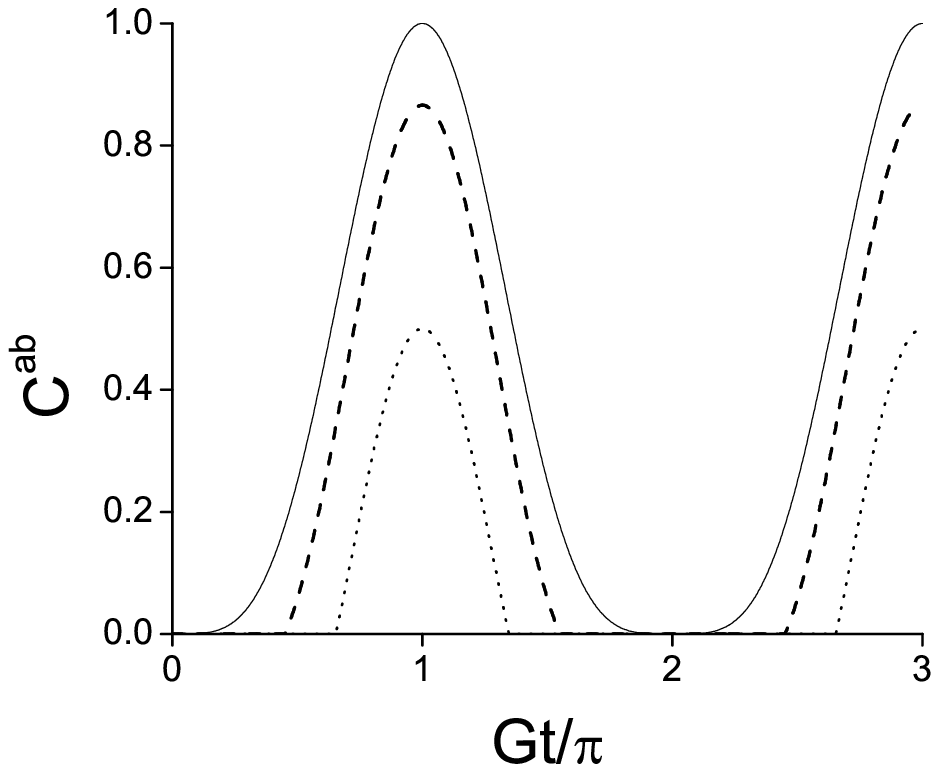, width=5cm}
\caption{{\footnotesize \label{fig3} Plots of atom-atom and cavity-cavity concurrences for three values of the superposition parameter $\alpha$. Entanglement sudden death occurs for all superpositions except the pure Bell states with $\alpha = (1\pm \frac{1}{2})\frac{\pi}{2}$.
}}
\end{figure}

Figs.~\ref{fig1}, \ref{fig2}, and \ref{fig3} show that $C^{AB}$ has a novel feature. The entanglement non-smoothly becomes and stays zero for a finite interval of time for a continuous range of $\alpha$ values.  This effect was first noted in \cite{Yu-Eberly04} and has been found in several contexts ( See, e.g., \cite{Yu-Eberly05,Ficek-Tanas06,Ban06,Lu,Jamroz06}). It has been referred to as ``entanglement sudden death" (ESD). Without commenting on the physical implications here, this is due simply to the fact that $Q^{AB}$ can take negative values. For $\al < \pi/4$ there is always a time interval in which entanglement remains zero while for $\pi/4 \leq \al \leq 3\pi/4$ entanglement becomes zero only momentarily (exactly at multiples of $t=\pi/G$).

From Eq.~(\ref{inverseJCstates}) it can be shown that for resonance,
after a time interval of $\pi/G$, the state $|e_A,0_a\rangle$ turns
into the state $|g_A,1_a\rangle$ up to an overall phase of $\pi/2$
and vice versa while the state $|g_A,0_a\rangle$ does not change.
The same is true for the $B$ and $b$ parts, of course.  Using this
fact it can be shown that the $|\Phi(t)\rangle$ of
Eq.~(\ref{mixedstate}) is the same as $|\Phi(t+\pi/G)\rangle$ under
an interchange between atomic and photonic states
($e_A\leftrightarrow 1_a, g_A\leftrightarrow 0_a, e_B\leftrightarrow
1_b, g_B\leftrightarrow 0_b$) up to a phase $\pi/2$ for resonance.
This indicates that the state seen from the perspective of atomic
part of the system at time $t$ is the same as the state that is seen
from the perspective of photonic part at time $t+\pi/G$ so the
concurrence between the cavities, $C^{ab}$, is the same as $C^{AB}$
up to a phase difference of $\pi/G$.  For the case of resonance, we
have \beq |z| - \sqrt{bc} = \frac{1}{4}\cos^2\alpha (2-2\cos G
t)[\tan\alpha - \cos^2 (G t/2)], \eeq so the expression for
concurrence turns out to be: \beq C^{ab} = 2\max\{0, Q^{ab} \}, \eeq
where $Q^{ab} = \cos^2\alpha\sin^2(Gt/2)[\tan\alpha -
\cos^2(Gt/2)]$. Thus a plot of $C^{ab}$ as a function of $\alpha$
and $t$ would look exactly like Fig.~\ref{fig1} except for a phase
shift by $\pi$ to the left on the time axis (see Fig.~\ref{fig3}).

Now we move to the concurrence calculations between opposite atoms and
cavities. That is,  we will calculate  $C^{Ab}$ and $C^{Ba}$ and $C^{Aa}$.

We start with $C^{Ab}$.  For this, we trace out $B$ and $a$ and find,
\beqa
\la g_B, 0_a\Phi(t)\ra &=&
\cos\alpha (cs_0 - sc_0)(cc_0 + ss_0)|e_A,1_b\ra\nonumber\\
&+&\sin\al|g_A,0_b\ra\nonumber\\
\la e_B,0_a|\Phi(t)\ra &=&
\cos\alpha (cc_0+ss_0)^2|e_A,0_b\ra\nonumber\\
\la g_B,1_a|\Phi(t)\ra &=&
\cos\al(cs_0-sc_0)^2|g_A,1_b\ra\nonumber\\
\la e_B,1_a|\Phi(t)\ra &=&
\cos\alpha (cs_0 - sc_0)(cc_0 + ss_0)|g_A,0_b\ra.\nonumber
\\
\eeqa
\beqa
|z| &=&|\sin\al\cos\al|c_0s_0\sqrt{2-2\cos \delta t}\nonumber\\
&\times & \sqrt{c_0^4 + s_0^4 +2c_0^2s_0^2\cos\delta t}\nonumber\\
b &=& \cos^2\al~(c_0^4 + s_0^4 +2c_0^2s_0^2\cos\delta t)^2\nonumber\\
c &=& \cos^2\al~ c_0^4s_0^4(2-2\cos \delta t)^2 .
\eeqa

For resonance
\beqa
\label{QAb}
Q^{Ab}&=&|z|-\sqrt{bc}\nonumber\\
&=&\frac{1}{4}\cos^2\al|\sin Gt|~(2|\tan \al|-|\sin Gt|).
\eeqa
Then the concurrence is
\beq C^{Ab}=2\max\{0, Q^{Ab}\}.
\eeq

Now if we look at the initial state in Eq.~(\ref{PZero}) we see that under the transformation $A\leftrightarrow B$,$a\leftrightarrow b$ the state remains unchanged.  That means $C^{Ba} = C^{Ab}$ at all times. Eq.~(\ref{QAb})  suggests that the maximum value $C^{Ab}$ can take is less than $1$  so the parts $A$ and $b$ cannot be maximally entangled unlike parts $A$ and $B$ (or $a$ and $b$).

\begin{figure}[!h]
\epsfig{file=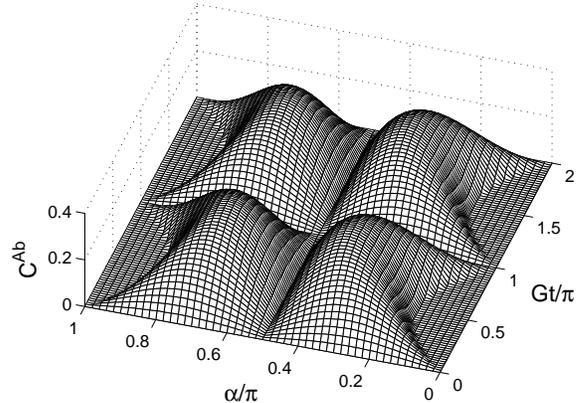, width=8cm} \caption{{\footnotesize
\label{fig5} Surface plot of the concurrence $C^{Ab}$ for the $|\Phi_{AB}\ra$ type of initial state as a function of $t$ and $\alpha$.
}}\end{figure}

\begin{figure}[!h]
\epsfig{file=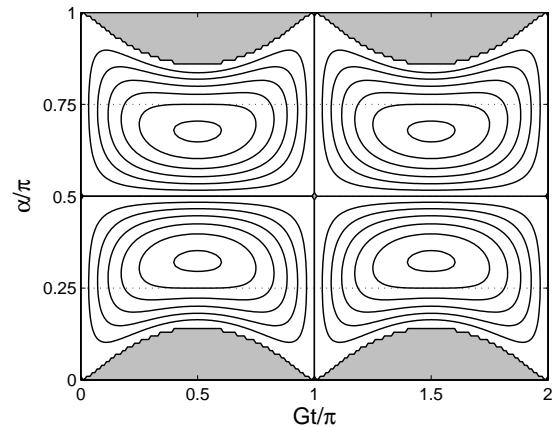, width=8.0 cm}
\caption{{\footnotesize \label{fig6}  Contour plot of the concurrence $C^{Ab}$ for the $|Phi_{AB}\ra$ type of initial state as a function of time and parameter $\alpha$. The innermost contours indicate a concurrence value of 0.3, and $\Delta_c = 0.05$ between contour lines. Regions of sudden death are painted in gray.}}
\end{figure}
As seen from plots (\ref{fig5}) and (\ref{fig6}),  there exist some parameter zones, where entanglement sudden death occurs for $C^{Ab}$ in a periodic manner. This interesting feature is not expected to be a generic feature for an arbitrary initial state, as shown in the next section.

Similarly, we can find the explicit expressions for $C^{Aa}$ and $C^{Bb}$.  The two cavities are not distinguishable, so we expect these $C$'s to be the same and they are. For resonance they become:
\beq
C^{Aa}=|\sin Gt|\cos^2\al = C^{Bb}.
\eeq

Before ending this section,  we want to stress that the onset of entanglement sudden
death (entanglement decays to zero in a finite time and remains zero for at least a finite time) is not due to the special mathematical structure of Wootters'  concurrence. As we have already mentioned, ESD simply signals the finite-time onset of separability (lack of entanglement), and this is a feature that will be confirmed by all entanglement measures.  More explicitly,
if a state at time $t_d$ is separable as measured by concurrence, then at the same time it is a separable
state by all other valid measures -- entanglement of formation, partial transpose negativity, etc.

\subsection{Partially entangled Bell States $|\Psi_{AB}\ra$}
\label{secondpair}

Now we examine entanglement dynamics for the following initial state:
\beqa
\label{PhiZero} |\Psi(0)\ra &=& |\Psi_{AB}\ra \otimes |0_a,0_b\ra\\
&=& (\cos\alpha|e_A, g_B\ra + \sin\alpha|g_A, e_B\ra) \otimes |0_a,
0_b\ra. \nonumber
\eeqa

Following the methods of the previous subsection \ref{firstpair}, we find:
\beqa
|\Psi(t)\ra &=& \cos\alpha \Big(c(t)|\psi_1^+\ra_A -
s(t)|\psi_1^-\ra_A\Big)\otimes |\psi^0\ra_B\nonumber\\
&+& \sin\alpha|\psi^0\ra_A \otimes\Big(c(t)|\psi_1^+\ra_B -
s(t)|\psi_1^-\ra_B\Big).
\eeqa

Changing into the bare basis:
\beqa
\label{phifunction}
|\Psi(t)\ra &=& \cos\alpha \Big(c(t)(c_0|e_A,0_a\ra + s_0|g_A,1_a\ra) \nonumber\\
&-& s(t)(-s_0|e_A,0_a\ra + c_0|g_A, 1_a\ra)\Big)
\otimes |g_B,0_B\ra \nonumber \\
&+& \sin\alpha|g_A, 0_a\ra \otimes\Big(c(t)(c_0|e_B,0_b\ra +
s_0|g_B,1_b\ra)\nonumber \\
&-& s(t)(-s_0|e_B,0_b\ra + c_0|g_B, 1_b\ra)\Big). \nonumber \\
\eeqa

Again, we start with $C^{AB}$. The projections that are needed are:
\beqa \label{abProjections}
\la 0_a, 0_b|\Psi(t)\ra &=& \cos\alpha(cc_0 + ss_0)|e_A,g_B\ra\nonumber\\
& +& \sin\alpha(cc_0 + ss_0) |g_A, e_B\ra \nonumber \\
\la 1_a,0_b|\Psi(t)\ra &=& \cos\alpha (cs_0 - sc_0)|g_A,g_B\ra \nonumber \\
\la 0_a,1_b|\Psi(t)\ra &=&\sin\alpha(cs_0-sc_0)|g_A,g_B\ra\nonumber\\
\la 1_a,1_b|\Psi(t)\ra &=&0.
\eeqa

The density matrix that these projections give is in the form of
Eq.~(\ref{mixedstate})  again.  Yet, this time we have only three non-zero diagonal entries instead of the four of the previous section.  This leads to a zero $\sqrt{bc}$ and makes it impossible to find sudden death since in this case the function $Q^{AB}$  cannot take negative values.

Continuing the calculations,
\beqa \label{absolutew}
|z|^2 &=& \sin^2\alpha\cos^2\alpha |(cc_0 + ss_0)|^4 \nonumber \\
&=& \sin^2\alpha\cos^2\alpha (c_0^4 + s_0^4 +2c_0^2s_0^2\cos\delta t)^2, \nonumber \\
|z| &=& |\sin\alpha \cos\alpha|(c_0^4 + s_0^4 +2c_0^2s_0^2\cos\delta t), \nonumber \\
b &=& \cos^2\alpha |cs_0 - sc_0|^2+\sin^2\alpha|cs_0 - sc_0|^2 \nonumber \\
&=& |cs_0 - sc_0|^2 \nonumber \\
&=&c_0^2s_0^2(2-2\cos\delta t),\nonumber\\
c &=& 0.
\eeqa

For resonance we find
\beq
Q^{AB}=|z| - \sqrt{bc} =
\frac{1}{4}|\sin\alpha\cos\alpha| (2+2\cos\delta t),
\eeq
where
$$\delta|_{\Delta =0} = G,$$
so the expression for concurrence turns out to be:
\beq
\label{ResConcurrence} C^{AB} = 2\max\{0, Q^{AB} \},
\eeq
where $Q^{AB} =|\sin\alpha\cos\alpha|\cos^2(Gt/2)$.  Since $Q\geq 0$, we get,
\beqa
 C^{AB}&=&2|\sin\alpha\cos\alpha|\cos^2(Gt/2)\nonumber\\
 &=&|\sin2\al|\cos^2(Gt/2).
 \eeqa

\begin{figure}[!h]
\epsfig{file=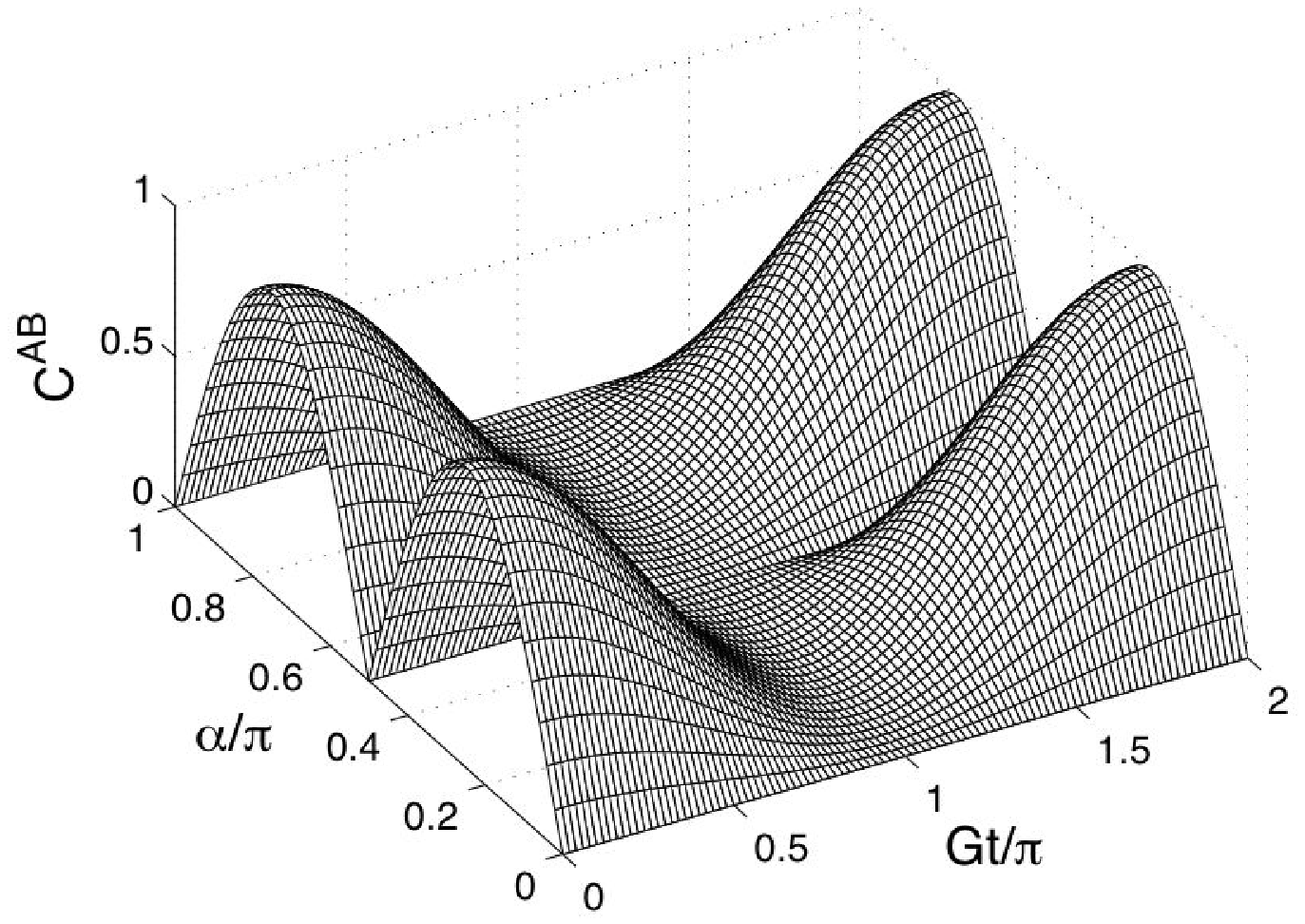, width=8cm} \caption{{\footnotesize
\label{fig7} Surface plot of the concurrence $C^{AB}$  for the $|\Psi_{AB}\ra$ type of initial state as a function of time and parameter $\alpha$.
}}\end{figure}

\begin{figure}[!h]
\epsfig{file=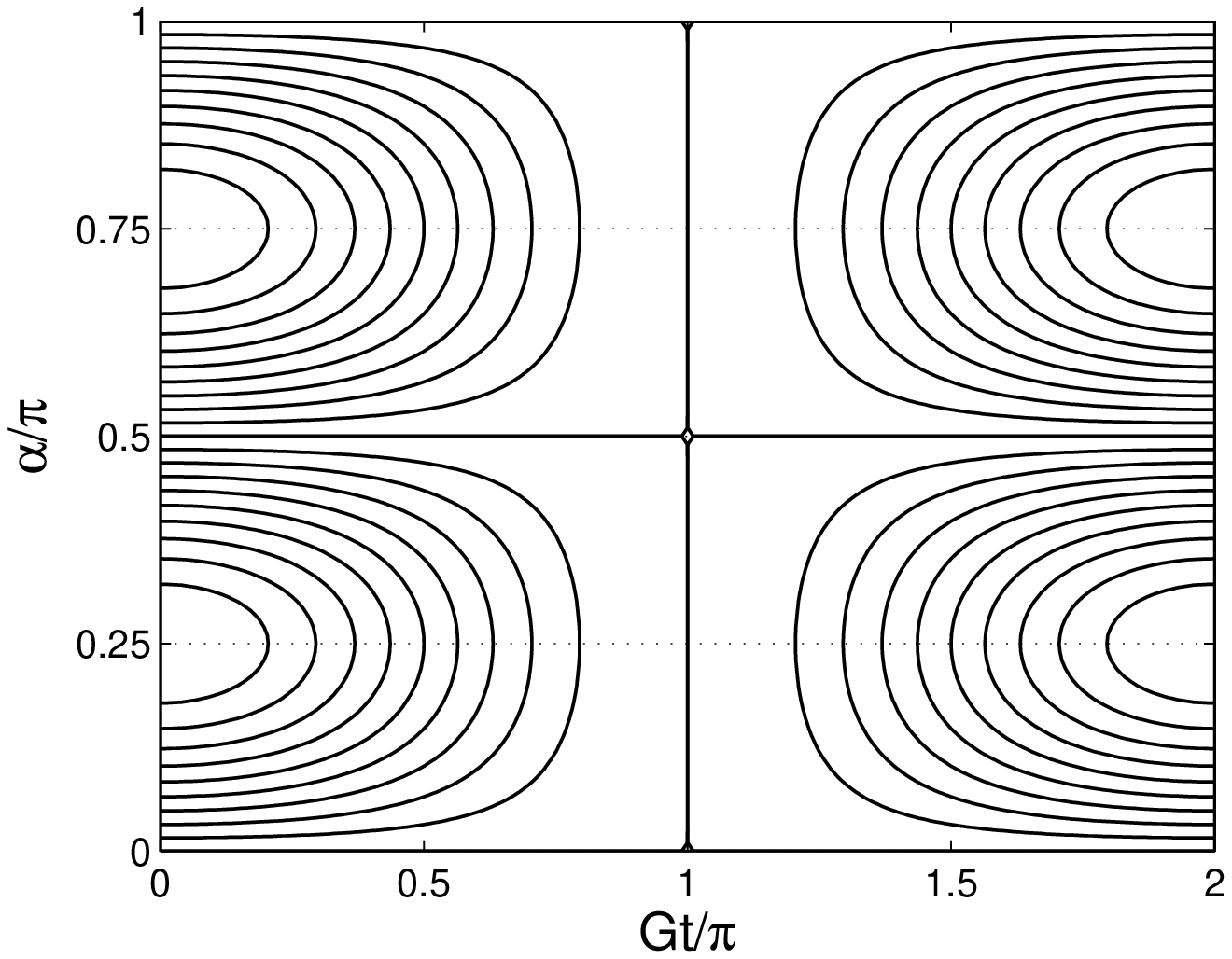, width=8.0 cm}
\caption{{\footnotesize \label{fig7a}  Contour plot of the
concurrence $C^{AB}$ for the second type of initial state as a
function of time and parameter $\alpha$. The innermost conturs indicate a concurrence value of 0.9, and the value drops by $\Delta_c = 0.1$ between two consecutive contours. Regions of sudden death are painted in
gray.}}
\end{figure}

\begin{figure}[!h]
\epsfig{file=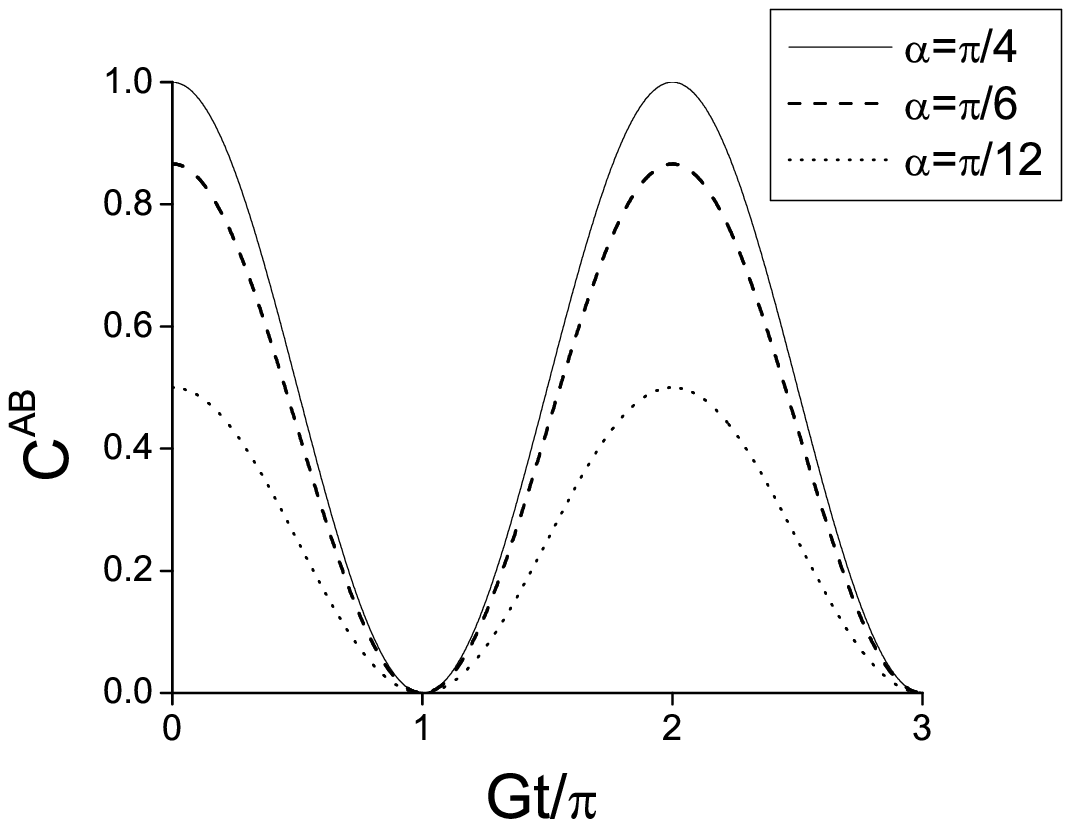, width=6cm}
\epsfig{file=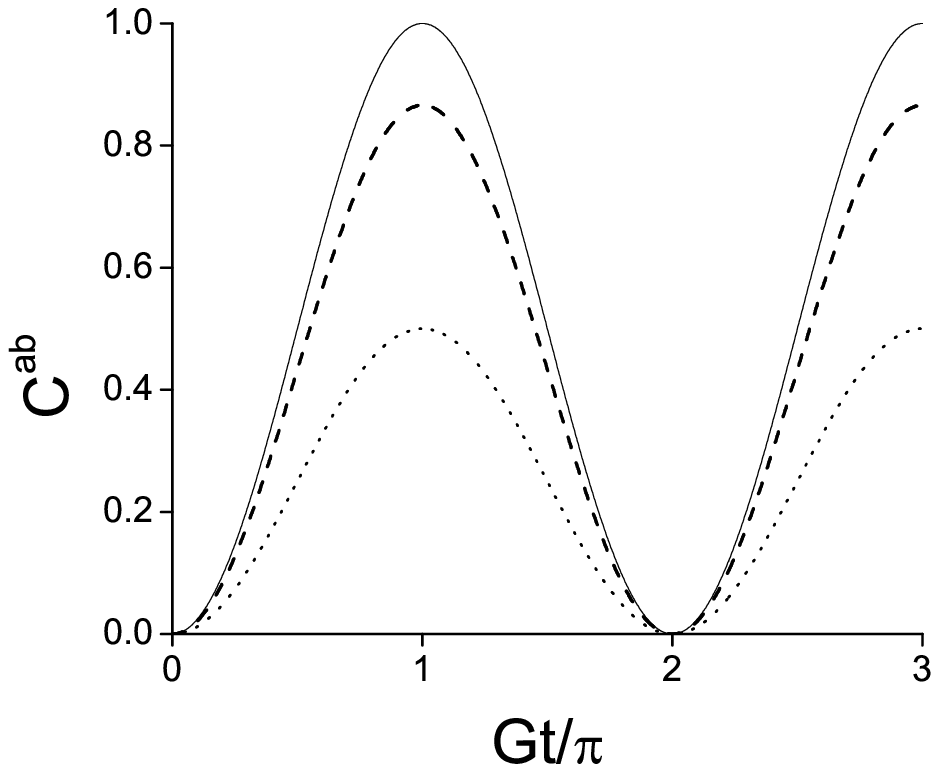, width=5cm} \caption{{\footnotesize
\label{fig9} Plots of atom-atom and cavity-cavity concurrences for three values of superposition parameter $\alpha$. Entanglement sudden death is now absent for all superpositions although $C=0$ is reached at a single time for the pure Bell states with $\alpha = (1\pm \frac{1}{2})\pi/2$.
}}
\end{figure}

From Figs.~(\ref{fig7}) and (\ref{fig7a}), as well as (\ref{fig9}),  we see that $C^{AB}$ (also $C^{ab}$) lacks the feature of
sudden death.  The plot  Fig.~\ref{fig7} touches the $x-$ axis only at multiples of $t=\pi/G$.

It can be shown that Eq.~(\ref{phifunction}) has the same symmetry as Eq.~(\ref{psifunction}) under
transformations $t\rightarrow t+\pi/G$,$e_A\leftrightarrow 1_a,
g_A\leftrightarrow 0_a, e_B\leftrightarrow 1_b$ and $
g_B\leftrightarrow 0_b$, so $C^{ab}(t+\pi/G)=C^{AB}(t)$ for
resonance. By considering symmetry between the atom and cavity,
again we have \beqa C^{ab}&=&2Q\nonumber\\
&=&|\sin2\al|\sin^2(Gt/2).\eeqa

The remaining concurrence between the atoms and cavities
are $C^{Ab}$ and $C^{Ba}$. If we trace out $B$ and $a$,
\beqa
\la g_B,0_a|\Psi(t)\ra &=&\cos\alpha(cc_0+ss_0)|e_A,0_b\ra\nonumber\\
&+& \sin\alpha (cs_0 - sc_0)|g_A,1_b\ra,\nonumber\\
\la e_B, 0_a|\Psi(t)\ra &=& \sin\alpha(cc_0 + ss_0)|g_A,0_b\ra,\nonumber\\
\la g_B,1_a|\Psi(t)\ra &=& \cos\alpha (cs_0 - sc_0)|g_A,0_b\ra, \nonumber \\
\la e_B,1_a|\Psi(t)\ra &=&0 . \eeqa
Again we only have three nonzero projections which means either $b$
or $c$ is zero.  In either case, calculating only the $|w|$ is
sufficient to find out the concurrence.

For the case of resonance, \beq
|z|=|\sin\al\cos\al|\frac{1}{4}\sqrt{2+2\cos G t}\sqrt{2-2\cos G
t}.\eeq Then the concurrence is \beqa \label{CAb}
 C^{Ab}&=&2|z|\nonumber\\
&=&2|\sin\al\cos\al||\sin Gt/2||\cos Gt/2|,\nonumber\\
 &=&|\sin\al\cos\al||\sin Gt|.\eeqa

Looking at Eq.~(\ref{PhiZero}) we see that under a transformation
$A\leftrightarrow B$,$a\leftrightarrow b$ and
$\cos\al\leftrightarrow\sin\al$ the state remains unchanged.  That
means if we just applied the transformation
$\cos\al\leftrightarrow\sin\al$ to Eq.~(\ref{CAb}) we would find $C^{Ba}$,
but such a transformation does not change this equation, so
$C^{Ba}=C^{Ab}$ at all times.

Eq.~(\ref{CAb}) suggests that the maximum value $C^{Ab}$ can take is $0.5$ so
the parts $A$ and $b$ cannot be maximally entangled unlike parts $A$
and $B$ (or $a$ and $b$).

In the exactly same way, for resonance we get
\beq
C^{Aa}=\cos^2\al|\sin Gt|.
\eeq
and
\beqa
C^{Bb}=\sin^2\al|\sin Gt|,
\eeqa
respectively.

\begin{figure}[!h]
\epsfig{file=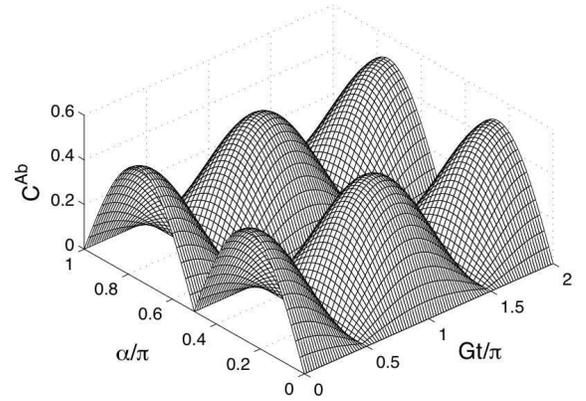, width=8cm} \caption{{\footnotesize
\label{fig21} Surface plot of the concurrence $C^{Ab}$ for the $|\Psi_{AB}\ra$ type of initial state as a function of time and parameter $\alpha$.
}}\end{figure}

\begin{figure}[!h]
\epsfig{file=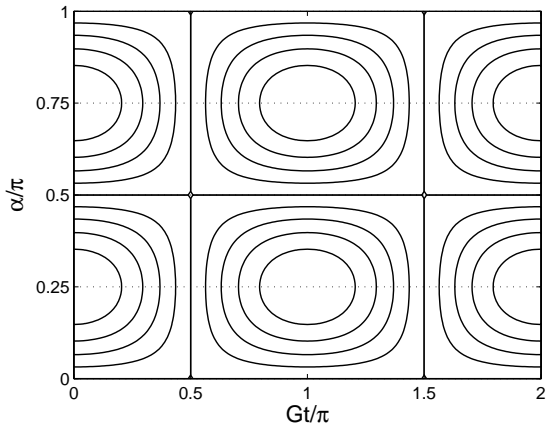, width=8.0 cm}
\caption{{\footnotesize \label{fig22}  Contour plot of the
concurrence $C^{Ab}$ for the $|\Psi_{AB}\ra$ type of initial state as a
function of time and parameter $\alpha$. The innermost loops has a
concurrence of 0.4 on them. Concurrence decreases linearly as we
move to the outer loops with $\Delta_c=0.1$.  No region of sudden death
is seen.}}\end{figure}

Similar to $C^{AB}$ for the initial state (\ref{PhiZero}), the plots  (\ref{fig21}) and
(\ref{fig22}) clearly show that  there is no sudden death feature for $C^{Ab}$.
From the calculations it is evident that $C^{AB}(t)$ and $C^{ab}(t)$
have only a phase difference of $Gt/\pi$ between them.  This can be
explained with excitation exchange between the the systems. That is
for instance from the Hamiltonian one can tell that what is lost
from $A$ in terms of excitation must go to $a$ so that after half a
Rabi cycle the two systems (A-B and a-b) exchange their initial
states. It is also interesting that for the second Bell state the
summation of $C^{AB}$ and $C^{ab}$ equals the initial concurrence
whereas for the first type of Bell state it is less than the initial
one.

For the $|\Phi_{AB}\ra$ type of initial state we obtain an extra entry in the density matrix ($|e_A,e_B\rangle\langle e_A,e_B|$ if we are looking
at the entanglement in the atomic part) that was not there for the
$|\Psi_{AB}\ra$ type. Existence of this leads to the negative term in the $Q$ function and thus to the sudden death. Without this negative term we
would expect to get the initial entanglement when we added up the
entanglements of the atomic and photonic parts of the system.
However this negative term causes a leak in the entanglement.

\section{Concluding comments}
\label{conclusion}

We have examined entanglement dynamics of several subsystems in a
four-qubit model in which two atoms are locally coupled to the
modes of their cavities.  A distinctive feature in this model is
that there are no couplings between the two atoms or the two cavities. That is, evolution of entanglement is a pure information-exchange evolution, not a more conventional decoherence process involving energy interaction and exchange.

We have studied the onset of entanglement sudden death for two sets of pure initial states. We demonstrate that there exist finite-time intervals in which the entanglement measured by concurrence remains zero.  Due to the absence of interactive decoherence in the pure JC model, the lost entanglement can bounce back in a finite time.  For the $|\Psi_{AB}\ra$ type of initial state, ESD does not occur. Although entanglement decays to zero in a finite time, it never remains zero. The occurrence of ESD for the $|\Phi_{AB}\ra$ state is not contradictory, and it is easy to see physical distinctions that make the two types of initial state nonsymmetric. In the $|\Phi_{AB}\ra$ initial case two photons may later be present at the same time, whereas there can never be more than one photon present in the $|\Psi_{AB}\ra$ case. This said, it is still unclear exactly what is the governing distinction. This remains an interesting open question in entanglement dynamics.  In addition, it should be pointed that our main results can be extended to the case that the cavity fields are not exactly in resonance with the  atoms. The qualitative features of our conclusions will remain the same, and the detailed dynamics in non-resonant situations will simply become more complicated.

In conclusion, we can make three comments: Firstly, it is not surprising that entanglement can be generated by couplings between atoms and photons.  What appears here is different. The local atom-cavity couplings not only cause the creation of entanglement between an atom and its cavity, but also generate  non-local entanglement between
the cavities if the atom-atom subsystem is initially prepared in an entangled state.
Moreover, as shown explicitly, with aid of entanglement between two atoms one may generate non-local  entanglement between one atom and a remote cavity mode.  Secondly, we have demonstrated  in the case of the $|\Psi_{AB}\ra$ type of initial state, the loss of entanglement $C^{AB}$ in the atom-atom system is instantly compensated by concurrence gain in $C^{ab}$ for the cavity-cavity system.  In fact,
it is easy to check that we have

\beq C^{AB} + C^{ab} = |\sin2\alpha| = {\rm const.} \eeq

We may interpret the entanglement gain in the cavity-pair as a
process of entanglement transfer from atomic variables to photonic
variables. However, it may also suggest some kind of  ``entanglement
conservation".  This is an open issue that is largely unexplored,
and we can't expect conservation of entanglement in the sense of
dynamical conservation laws, since entanglement is not defined as an
observable or represented by a Hermitian operator. Interestingly, we
can follow the negative values present in the concurrence with the
aid of the parameter $Q^{AB}$.  The following equation  holds for
all the initial states (\ref{PZero}) and  (\ref{PhiZero}) considered
in this paper: \beq Q^{AB}+Q^{ab}+2Q^{Aa}|\tan\al|-2Q^{Ab}=|\sin
2\al|, \eeq where the right hand side is nothing but the initial
concurrence. This is an issue to be explored elsewhere.  Finally,
our model may suggest an interesting link between a system's
temporal memory and entanglement generation.  In the case of the JC
model, the memory effect is due to the fact that the  photons
absorbed by the cavity modes will be coupled back to the atom in a
finite time.  In a more general context, our research suggests that
a non-Markovian property may play a crucial role in recovering
entanglement in a quantum dynamical system.

\section*{Acknowledgements}

\noindent

We acknowledge financial support from NSF Grant PHY-0456952 and ARO Grant W911NF-05-1-0543.



\begin{thebibliography}{99}

\bibitem{MC} M.\ A.\ Nielsen and I.\ L.\ Chuang, {\it Quantum
Computation and Quantum Information} (Cambridge Univ. Press, 2000).

\bibitem{Slichter78} See, for example, C.\ P.\  Slichter,   {\it Principles of Magnetic
Resonance} (Springer-Verlag, Berlin, 1978), 2nd ed.

 \bibitem{Yu-Eberly02} T. \ Yu and J.\ H.\  Eberly, \prb{66}, 193306
(2002).

\bibitem{Yu-Eberly03} T. \ Yu and J.\ H.\  Eberly, \prb{68}, 165322 (2003).

\bibitem{Yu-Eberly04} T. \ Yu and J.\  H.\  Eberly, \prl {93}, 140404 (2004).

\bibitem{Jakobczyk-Jamroz04} L. \  Jakobczyk and A.\   Jamroz,  Phys.  Lett.
A  {\bf 333}, 35 (2004).

\bibitem{Lidar04} S. Bandyopadhyay and D. A. Lidar, Phys. Rev. A 70, 010301(R) (2004).

\bibitem{Tolkunov-etal05} D. \ Tolkunov, V.\   Privman, and P.\  K.  \
Aravind, Phys. Rev. A {\bf 71}, 060308 (2005).


\bibitem{Yu-Eberly05} T. \ Yu and J.\ H.\  Eberly,  Opt. Comm.  {\bf 264},  393 (2006).
and arXiv: quant-ph/0503089.

\bibitem{Yu-Eberly06B} T. \ Yu and J.\ H.\  Eberly,  Phys.  Rev. Lett.  {\bf 97}, 140403 (2006).


\bibitem{Ficek-Tanas06} Z. \ Ficek and R. \ Tanas,  Phys. Rev. A {\bf 74}, 024304 (2006).

\bibitem{Halliwell} P.\ J.\  Dodd and J.\ J.\  Halliwell, \pra {69},
052105 (2004).

\bibitem{Ban06} M.\ Ban, J. Phys. A  {\bf 39}, 1927 (2006).

\bibitem{Solenov05} D.\  Solenov, D.\ Tolkunov and V.\  Privman,  Phys. Lett.  {\bf 359}, 81 (2006).


\bibitem{Zyczkowski02} K. Zyczkowski, P.\ Horodecki, M. \ Horodecki
and R.  Horodecki, \pra{65}, 012101 (2002).

\bibitem{Diosi03} L.\  Diosi, in {\em Irreversible Quantum Dynamics},
edited by F. \ Benatti and R.\  Floreanini (Springer, New York, 2003),
pp. 157-163.

\bibitem{Carvalho-etal04}  A.\ R.\ R. Carvalho, F.\  Mintert and A.\
Buchleitner, \prl {93}, 230501 (2004).

\bibitem{Carvalho-etal05} A.\ R.\ R.\   Carvalho, F.\  Mintert, S.\  Palzer and
A.\  Buchleitner,  Euro.   Phys.  J.  D 41, 425 (2007).


\bibitem{phys} F.\  Mintert, A.\ R. \ R. \ Carvalho, M.\ Kus, A.   Buchleitner, Phys. Rep.  {\bf  415}, 4,  207(2005).

\bibitem{Lu} L.\  Derkacz and L.\  Jakobczyk,  \pra {74},   032313  (2006).

\bibitem{kuzi0} G.\  Gordon, G. \ Kurizki, and A.\  G. \ Kofman, J. Opt. B,  7, S283 (2005).

\bibitem{kuzi1} G. Gordon and G.\ Kurizki,  \prl{97}, 110503 (2006);  G. Gordon and G.\ Kurizki,  submitted to J. Phys. B  (2007).

\bibitem{kuri2}  L.\ Viola, E.\ Knill, and S.\  Lloyd, Phys. Rev. Lett. {\bf 85}, 3520 (2000);
L.\ A.\ Wu and D.\ A.\ Lidar, Phys. Rev. Lett. {\bf 88}, 207902 (2002);
 E.\ Knill, R.\ Laflamme, and L.\ Viola, Phys. Rev. Lett. {\bf 84}, 2525 (2000);
 P.\  Zanardi and S.\ Lloyd, Phys. Rev. Lett. {\bf 90}, 067902 (2003).


\bibitem{sun}  Z.\ Sun, X.  Wang, Y.\ B.\ Gao, C.\ P.\ Sun, quant-ph/0701093.

\bibitem{Santos-etal06} M. \ F.\  Santos, P.\  Milman, L.\  Davidovich  et al, \pra{73},  022313 (2006).

\bibitem{nature}S. \ P.\  Walborn,  P.\  H.\  S.\  Ribeiro, L.\  Davidovich, F.\  Mintert, A. \ Buchleitner,  Nature, {\bf 440},  1022(2006).

\bibitem{Davi}M.\ P.\ Almeida, F.\ de Melo, A.\ Salles, M.\  Hor-Meyll, S.\ P.\ Walborn, P.\ H.\ Souto Ribeiro, and L.\ Davidovich (Quantum Optics III  Conference, Pucon, Chile, 2006).

\bibitem{Yonac} M.\ Y\"{o}na\c{c}, T.\  Yu, and J.\ H.\  Eberly, \jpb {39},   S621 ( 2006).


\bibitem{JC} E.\ T.\  Jaynes and F.\ W.\  Cummings, Proc. IEEE {\bf 51}, 89 (1963).

\bibitem{Eberlyetal80} J.\  H. \ Eberly,  N. \ B.\  Narozhny, and J.\  J. \ Sanchez-Mondragon,
Phys. Rev. Lett. {\bf 44}, 1323 (1980).

\bibitem{Harochegroup} M. Brune, F. Schmidt-Kaler, A. Maali, J. Dreyer, E. Hagley, J. M. Raimond, and S. Haroche, Phys. Rev. Lett. {\bf 76}, 1800 (1996).


\bibitem{Winelandgroup} D. M. Meekhof, C. Monroe, B. E. King, W. M. Itano, and D. J. Wineland, Phys. Rev. Lett. {\bf 76}, 1796 (1996).


\bibitem{Kimblegroup}  A. Boca, R. Miller, K. M. Birnbaum, A. D. Boozer, J. McKeever, and H. J. Kimble, Phys. Rev. Lett. {\bf 93}, 233603 (2004)


\bibitem{Banacloche} J.\  Gea-Banacloche, \prl{65}, 3385 (1990).


\bibitem{Wootters} See  W.\ K.\  Wootters, \prl{80},  2245 (1998); S. \
Hill and W.\ K. \ Wootters, Phys. Rev. Lett. {\bf 78}, 5022 (1997).


\bibitem{ye1} T. \ Yu and J. \ H. \ Eberly, Quantum Inf. and Comp. (in press, 2007) and e-print quant-ph/0503089.

\bibitem{Jamroz06} A. Jamroz, J.  Phys.  A  {\bf 39}, 7727 (2006).


\end{thebibliography}
\end{document}